\def\caln{{\cal N}}
\def\cald{{\cal D}}
\def\DM{D$^-$}
\def\DP{D$^+$}
\def\NSM{NS$^-$}
\def\NSP{NS$^+$}
\def\HSM{HS$^-$}
\def\HSP{HS$^+$}
\def\etc{{\it etc.}}
\def\leff{L_{\rm eff}}
\def\sign{{\rm sign}}
\def\br{B}
\def\breff{\br_{\rm eff}}

\def\rts{\sqrt s}
\def\eps{\epsilon}

\def\epem{e^+e^-}
\def\mupmum{\mu^+\mu^-}

\def\taup{\tau^+}
\def\taum{\tau^-}

\def\lsim{\mathrel{\raise.3ex\hbox{$<$\kern-.75em\lower1ex\hbox{$\sim$}}}}
\def\gsim{\mathrel{\raise.3ex\hbox{$>$\kern-.75em\lower1ex\hbox{$\sim$}}}}
\def\@versim#1#2{\vcenter{\offinterlineskip
        \ialign{$\m@th#1\hfil##\hfil$\crcr#2\crcr\sim\crcr } }}

\def\slash#1{#1\hskip-6pt/\hskip2pt}

\def\etmiss{\slash E_T}

\def\ie{{\it i.e.}}

\def\anti{\overline}

\def\fbi{~{\rm fb}^{-1}}

\def\gev{\,{\rm GeV}}
\def\tev{\,{\rm TeV}}

\def\wt{\widetilde}

\def\mhalf{m_{1/2}}

\def\slep{\wt \ell}

\def\slepr{\wt \ell_R}
\def\mslepr{m_{\slepr}}

\def\hl{h^0}
\def\hh{H^0}
\def\ha{A^0}
\def\hp{H^+}
\def\hm{H^-}
\def\hpm{H^{\pm}}

\def\mhh{m_{\hh}}
\def\mha{m_{\ha}}
\def\mhp{m_{\hp}}
\def\mhpm{m_{\hpm}}
\def\tanb{\tan\beta}

\def\mt{m_t}

\def\mz{m_Z}

\def\wp{W^+}
\def\wm{W^-}

\def\cnone{\wt\chi^0_1}
\def\cntwo{\wt\chi^0_2}
\def\snu{\wt\nu}

\def\mcnone{m_{\cnone}}
\def\mcntwo{m_{\cntwo}}

\def\cpone{\wt \chi^+_1}
\def\cmone{\wt \chi^-_1}
\def\cpmone{\wt \chi^{\pm}_1}
\def\mcpone{m_{\cpone}}
\def\mcpmone{m_{\cpmone}}

\documentclass[smus]{snow2e}
\input psfig.sty
\begin{document}

\title{
{\normalsize \hspace*{\fill} UCD-96-40 \\
 \hspace*{\fill} October, 1996 \\}
Using Higgs Pair Production in {\boldmath $\epem$} or {\boldmath $\mupmum$}
Collisions To Probe GUT-Scale Boundary Conditions in the Minimal
Supersymmetric Model\thanks{
To appear in ``Proceedings of the 1996 DPF/DPB Summer Study
on New Directions for High Energy Physics''.
Work supported in part by the Department of Energy
and in part by the Davis Institute for High Energy Physics.}}

\author{J.~F. Gunion and J. Kelly\thanks{Current
address: Dept. of Physics, University of Wisconsin, Wisconsin, Madison WI
53706.} \\
{\it Davis Institute for High Energy Physics, University
of California, Davis, California 95616} 
}

\maketitle

\thispagestyle{empty}\pagestyle{plain}

\begin{abstract}
We delineate the techniques and prospects for using
Higgs pair production in $\epem$ or $\mupmum$ collisions
to probe GUT-scale boundary conditions in the minimal supersymmetric
standard model.
\end{abstract}

\section{Introduction and Results}

The heavier CP-even, the CP-odd and the charged
Higgs bosons ($\hh$, $\ha$, and $\hpm$, respectively)
of the minimal supersymmetric standard model (MSSM) 
(see Ref.~\cite{dpfreport} for a recent review)
are typically predicted to be fairly heavy (with
$\mhh\sim\mha\sim\mhpm\gsim 200\gev$) in models
where electroweak symmetry breaking at scale $\mz$
arises as a result of evolution
beginning from simple GUT/Planck-scale boundary conditions.
Nonetheless, once $\rts$ is large enough that $\epem\to \hh\ha$ and
$\epem\to \hp\hm$ (or their $\mupmum$ analgoues) are kinematically
possible, event rates are substantial for expected machine luminosities,
and discovery and study of these Higgs bosons becomes possible
\cite{pairpaper}. 
The all-jet and high-multiplicity
final states coming from $\hh,\ha\to b\anti b,t\anti t$
and $\hp\to t\anti b,\hm\to b\anti t$ are background free and
for the model we study provide appropriate and efficient signals
with rates that are adequate even when SUSY decays are present. 
Further, in the all-jet channels, the individual
Higgs boson masses, $\mha$, $\mhh$ and $\mhp$, can be measured.
Event rates and decay branching
fractions are typically such that it will be possible to `tag'
one member the pair in such a fully reconstructable final state
and then study the decays of the untagged member of the
pair. Here, we point out the very dramatic sensitivity of measurements of
decay branching fractions to the GUT boundary condition scenario,
illustrating in particular the high statistical level at which various not
terribly different scenarios can be distinguished from one another
using ratios of branching fractions. A more detailed treatment
of this analysis appears in Ref.~\cite{pairpaper}.

In the simplest GUT treatments of the MSSM, soft supersymmetry breaking
at the GUT scale is specified by three universal parameters:
\begin{itemize}
\item $m_0$: the universal soft scalar mass;
\item $\mhalf$: the universal soft gaugino mass;
\item $A_0$: the universal soft Yukawa coefficient.
\end{itemize}
The absolute value of $\mu$ (the Higgs mixing parameter) is 
determined by requiring
that radiative EWSB gives the exact value of $\mz$
for the experimentally measured value of $\mt$; however,
the sign of $\mu$ remains undetermined. Thus,
the remaining parameters required to completely fix the model are
\begin{itemize}
\item $\tanb$: the vacuum expectation value ratio; and
\item sign($\mu$).
\end{itemize}
We remind the reader that a universal gaugino mass at the GUT scale
implies that $M_3:M_2:M_1\sim 3:1:1/2$ at scale $\sim\mz$. For models
of this class one also finds that $\mu\gg M_{1,2}$. These two facts
imply that the $\cnone$ is mainly bino, while $\cntwo$ and $\cpone$
are mainly wino, with heavier gauginos being mainly higgsino, so that
$\mcntwo\sim\mcpone\sim 2\mcnone$.

We will consider three representative GUT scenarios characterized by
increasingly large values of $m_0$ relative to $\mhalf$ (which
translates into increasingly large slepton masses as compared to
$\mcnone$, $\mcntwo$, and $\mcpone$):
\begin{itemize}
\item ``No-Scale'' (NS): $A_0=m_0=0$;
\item ``Dilaton'' (D): $\mhalf=-A_0=\sqrt3 m_0$;
\item ``Heavy-Scalar'' (HS): $m_0=\mhalf$, $A_0=0$.
\end{itemize}
Within any one of these three scenarios, the model is completely specified
by values for $\mhalf$, $\tanb$ and sign($\mu$). We will present results
in the $(\mhalf,\tanb)$ parameter space for a given sign($\mu$) and
a given choice of scenario. Our notation will be $NS^-$ for the no-scale
scenario with $\sign(\mu)<0$, and so forth.

In exploring each of these scenarios, we proceed as follows.
\begin{itemize}
\item 
First, we delineate the allowed region of $(\mhalf,\tanb)$
parameter space consistent with all available experimental
and phenomenological constraints (such as the LSP being uncharged,
coupling constants remaining perturbative, no Higgs or SUSY
particle having been observed at LEP, \etc). The extent of these
regions is quite limited for the NS models, and is largest
for the HS models.
\item
Second, we determine the masses of the Higgs bosons and SUSY particles as
a function of $(\mhalf,\tanb)$.
The masses of the inos and the sleptons will presumably be
measured quite accurately, and 
they will determine the values of $\mhalf$ and $m_0$; but 
$\tanb$ is likely to be poorly determined
from these masses alone. However, 
a measurement of $\mha$ (or $\mhh$ or $\mhpm$) in combination with
the $\mhalf$ determination from the ino masses
will fix a value of $\tanb$.  The accuracy of this determination depends
upon the accuracy with which $\mha$ (and $\mhh$, $\mhpm$) can be measured.
For the $\ha,\hh\to b\anti b$ decay modes, for example,
this accuracy is fixed by the $b\anti b$ mass resolution.  
A resolution of $\pm\Delta M_{bb}\sim \pm10\gev$ 
is probably attainable.  For a 
large number, $N$, of events, $\mha$ can be fixed to a value of order
$\Delta M_{bb}/\sqrt N$, which for $N=20$ (our minimal
discovery criterion) would imply $\Delta \mha\sim 2-3\gev$.
Such mass uncertainty will lead to a rather
precise $\tanb$ determination within a given GUT model (except
in special cases).
\item
Finally, we examine the Higgs branching ratios as a function
of location in $(\mhalf,\tanb)$ parameter space, and determine
the statistical accuracy with which these branching ratios can
be measured for reasonable assumptions regarding Higgs tagging
and reconstruction efficiencies.
\end{itemize}

Ratios of branching ratios are of particular interest since
certain types of sytematic errors will cancel.  Relative Higgs branching
ratios can be measured by `tagging' one member of the produced
pair using a fully reconstructable all-jet decay mode, and then
looking at the various final states emerging from the decay
of the other member of the pair. Using the measured
values of $\br(\hl\to b\anti b)$ and $\br(t\to 2j b)$ and with
experimental knowledge of efficiencies, we can thus measure
\begin{eqnarray}
&{\br(\hh\to {\rm SUSY})\breff(\ha\to b\anti b+t\anti t)+\br(\ha\to 
{\rm SUSY})\breff(\hh\to
b\anti b+t\anti t) \over \breff(\hh\to b\anti b+t\anti t) 
\breff(\ha\to b\anti b+t\anti t)}
&\label{hhhasusy} \\
&{\br(\hh\to t\anti t )\br(\ha\to b\anti b)+\br(\ha\to t\anti t)\br(\hh\to
b\anti b) \over \br(\hh\to b\anti b) \br(\ha\to b\anti b)}
&\label{hhhattbb} \\
&{\br(\hh\to \hl\hl)\br(\ha\to b\anti b)\over \br(\hh\to b\anti b)
\br(\ha\to b\anti b)}
&\label{hhhlhl}\\
&{\br(\ha\to Z\hl)\br(\hh\to b\anti b)\over \br(\hh\to b\anti b)
\br(\ha\to b\anti b)}
&\label{hazhl}\\
&{\br(\hp \to {\rm SUSY})\br(\hm\to b\anti t)+\br(\hm\to {\rm SUSY})\br(\hp\to
t\anti b)\over  \br(\hp\to t\anti b)\br(\hm\to b\anti t)}
&\label{hphmsusy}\\
&{\br(\hp \to {\taup \nu})\br(\hm\to b\anti t)+\br(\hm\to {\taum \nu})\br(\hp\to
t\anti b)\over  \br(\hp\to t\anti b)\br(\hm\to b\anti t)}
&\label{hptaunu}\\
&{\br(\hp \to {\hl\wp})\br(\hm\to b\anti t)+\br(\hm\to {\hl\wm})\br(\hp\to
t\anti b)\over  \br(\hp\to t\anti b)\br(\hm\to b\anti t)}\,.
&\label{hpwhl}
\end{eqnarray}
The SUSY final states can be identified by the presence of missing
energy opposite the fully reconstructable all jet mode(s) used
to tag the first member of the Higgs pair.
We retain both $b\anti b$ and $t\anti t$ final states in
Eq.~(\ref{hhhasusy}), using an efficiency weighted combination 
denoted by $\breff$, in order that we may assess the importance
of SUSY decays both in regions where $b\anti b$ decays of
the $\hh,\ha$ are dominant and in regions where $t\anti t$ decays
are important. Note that since $\mha\sim\mhh$ we cannot separate the
$\hh$ and $\ha$ decays to the same final state; we can only measure the indicated
`average' values of Eqs.~(\ref{hhhasusy}) and (\ref{hhhattbb}).

\begin{figure}[[htb]
\leavevmode
\begin{center}
\centerline{\psfig{file=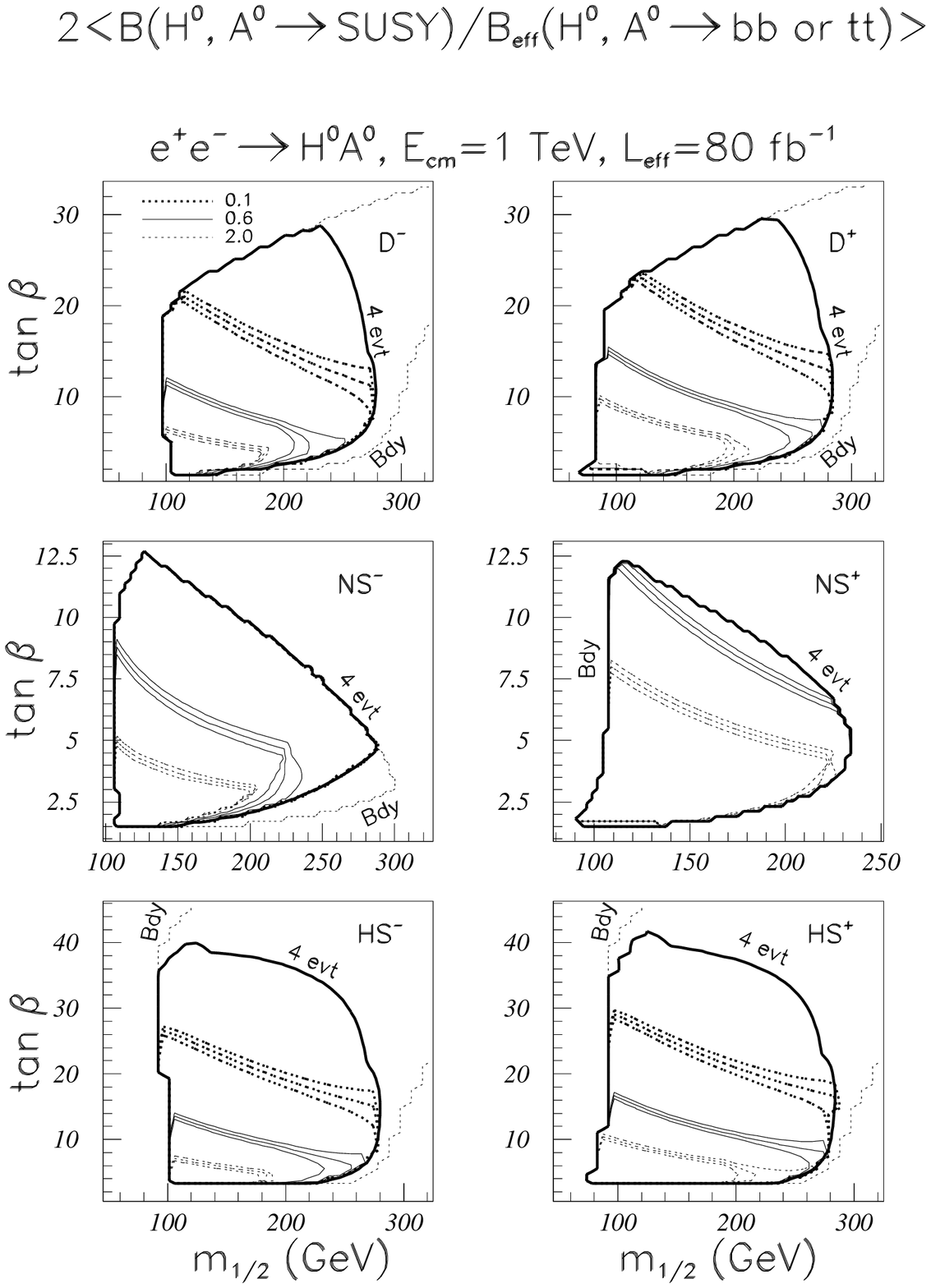,width=3.5in}}
\end{center}
\caption{
We plot contours along which the ratio 
of Eq.~(\ref{hhhasusy})
has a given constant value,
within the constraint/kinematically allowed
$(\mhalf,\tanb)$ parameter space (as indicated by the `Bdy' lines)
of the \DM, \DP, \NSM, \NSP, \HSM, and
\HSP\ models. Results are shown for the same three central values
for all models. For each central value, three lines are drawn. The central
line is for the central value. The other two lines 
are contours for which the ratio deviates by $\pm1\sigma$ statistical
error (see Ref.~\protect\cite{pairpaper}) from the central value.
Bold lines indicate the boundary beyond which fewer
than 4 events are found in the final states used to measure the numerator
of the ratio.}
\label{fhhhasusy}
\end{figure}

\begin{figure}[[htb]
\leavevmode
\begin{center}
\centerline{\psfig{file=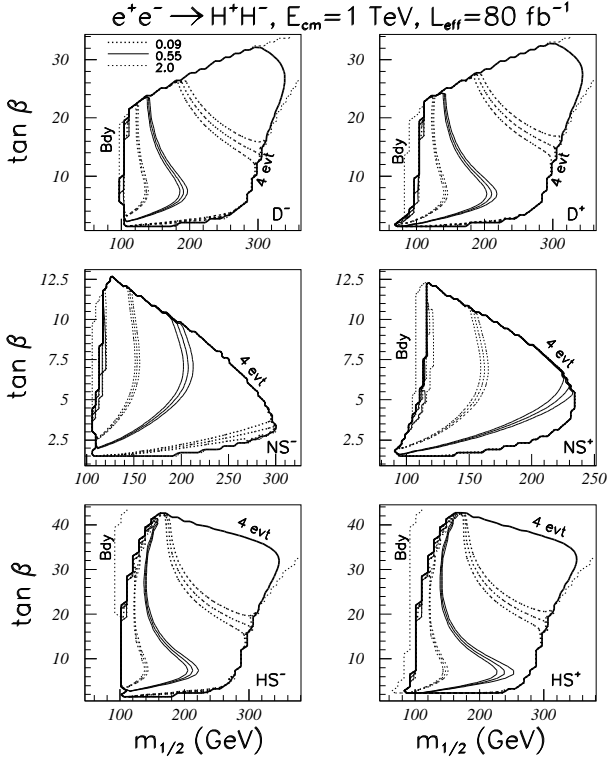,width=3.5in}}
\end{center}
\caption{
As in Fig.~\ref{fhhhasusy}, but for the ratio of Eq.~(\ref{hphmsusy}).}
\label{fhphmsusy}
\end{figure}

Two illustrations are provided. In Figs.~\ref{fhhhasusy} 
and \ref{fhphmsusy} we show
contours in $(\mhalf,\tanb)$ parameter space
of constant values for the ratio of Eqs.~(\ref{hhhasusy})
and (\ref{hphmsusy}), respectively, for each of the six scenarios defined
earlier. The three different curves for each value of the ratio
indicate the precision with which experiment can determine a location
in parameter space.  These results are based on event rates
calculated including all relevant branching ratios and assuming
an `effective' integrated luminosity of $L_{\rm eff}=80\fbi$ at $\rts=1\tev$, 
where $L_{\rm eff}=80\fbi$ includes an overall tagging, detection, 
and so forth, efficiency of $\eps=0.2$ at $L=400\fbi$ (about two years
of running). We observe that the precision is actually rather good.
Since the $\hh\ha$ and $\hp\hm$ SUSY ratio contours displayed tend
to cross one another, a measurment of these two ratios will
determine a location in $(\mhalf,\tanb)$ parameter space in each
GUT scenario.  It turns out that this determination 
in one model often disagrees
at a statistically very significant level
with the location determined on the basis of the $\mcpmone$
and $\mha$ masses, described earlier, for any other model.

To more thoroughly illustrate the extent to which the set of ratios given in
Eqs.~(\ref{hhhasusy})-(\ref{hpwhl}) can distinguish between scenarios,
let us focus on one particular case.
Suppose the correct model is \DM\ with $\mhalf=201.7\gev$ and 
$\tanb=7.50$. This would imply $\mha=349.7\gev$, $\mcpmone =149.5\gev$.
The $\mhalf$ and $\tanb$ values required in order
to reproduce these same $\mha$ and $\mcpmone$ values
in the other scenarios are listed in Table~\ref{mhalftanbtable}.
Also given in this table are the predicted values of $\mhh$
and $\mslepr$ for each scenario.
In order to get a first feeling for event numbers and 
for the errors that might be expected for the ratios of interest, we give in
Table~\ref{ratestable} the numbers of events, $\caln$ and $\cald$, predicted
in each scenario for use in determining
the numerators and denominators of Eqs.~(\ref{hhhasusy})-(\ref{hazhl})
and Eqs.~(\ref{hphmsusy})-(\ref{hpwhl}), assuming
$\leff=80\fbi$ at $\rts=1\tev$. These
numbers include the SUSY branching fractions, $\breff$, and so forth.

\begin{table}[hbt]
\caption[fake]{We tabulate the values of $\mhalf$ (in GeV)
and $\tanb$ required in each of our six scenarios in order
that $\mha=349.7\gev$ and $\mcpmone=149.5\gev$.
Also given are the corresponding values of $\mhh$
and $\mslepr$. Masses are in GeV.}
\begin{center}
\begin{tabular}{|c|c|c|c|c|c|c|}
\hline
 & \DM\ & \DP\ & \NSM\ & \NSP\ & \HSM\ & \HSP\ \\
\hline
\hline
$\mhalf$ & 201.7 & 174.4 & 210.6 & 168.2 & 203.9 & 180.0 \\
$\tanb$ & 7.50 & 2.94 & 3.24 & 2.04 & 12.06 & 3.83 \\
$\mhh$ & 350.3 & 355.8 & 353.9 & 359.0 & 350.1 & 353.2 \\
$\mslepr$ & 146.7 & 127.5 & 91.0 & 73.9 & 222.9 & 197.4 \\
\hline
\end{tabular}
\end{center}
\label{mhalftanbtable}
\end{table}

\begin{table}[hbt]
\caption[fake]{We give the numbers of events predicted
in each scenario at the parameter
space locations specified in Table~\ref{mhalftanbtable}
available for determining the numerators and denominators of 
Eqs.~(\ref{hhhasusy})-(\ref{hazhl}) and Eqs.~(\ref{hphmsusy})-(\ref{hpwhl}).
These event rates are those for $\leff=80\fbi$ at $\rts=1\tev$. 
They include all branching fractions.
Our notation is $\caln_{(\#)}$ and $\cald_{(\#)}$ 
for the event rates in the numerator and denominator, respectively,
of the ratio defined in Eq.~(\#).
}
\begin{center}
\begin{tabular}{|c|c|c|c|c|c|c|}
\hline
 & \DM\ & \DP\ & \NSM\ & \NSP\ & \HSM\ & \HSP\ \\
\hline
\hline
$\caln_{(\ref{hhhasusy})}$ & 97.0 & 92.3 & 88.3 & 49.2 & 76.1 & 124.0 \\
$\caln_{(\ref{hhhattbb})}$ & 0.1 & 0.7 & 3.8 & 1.02 & 0.0 & 0.2 \\
$\caln_{(\ref{hhhlhl})}$   & 16.4 & 2.7 & 46.6 & 1.47 & 3.8 & 2.4 \\
$\caln_{(\ref{hazhl})}$ & 2.0 & 1.3 & 9.2 & 0.6 & 0.4 & 1.1 \\
$\cald_{(\ref{hhhasusy})}$ 
      & 198 & 9.6 & 62.1 & 2.6 & 250 & 18.2 \\
$\cald_{(\ref{hhhattbb})-(\ref{hazhl})}$ 
      & 198 & 8.9 & 58.3 & 1.6 & 250 & 18.0 \\
$\caln_{(\ref{hphmsusy})}$ & 225 & 189 & 138 & 135 & 189 & 262 \\
$\caln_{(\ref{hptaunu})}$ & 58.4 & 4.2 & 6.5 & 1.1 & 90.0 & 9.5 \\
$\caln_{(\ref{hpwhl})}$ & 13.0 & 12.8 & 21.9 & 9.0 & 3.3 & 12.3 \\
$\cald_{(\ref{hphmsusy})-(\ref{hpwhl})}$ 
      & 317 & 415 & 445 & 465 & 320 & 348 \\
\hline
\end{tabular}
\end{center}
\label{ratestable}
\end{table}

\begin{table}[hbt]
\caption[fake]{We tabulate $\Delta\chi^2_i$ 
(relative to the \DM\ scenario) for the indicated branching
fraction ratios as a function of scenario,
assuming the measured $\mha$ and $\mcpmone$ values are $349.7\gev$
and $149.5\gev$, respectively. The SUSY channels have been resolved into 
final states involving a fixed number of leptons.  
The error used in calculating each $\Delta\chi^2_i$ is the approximate
$1\sigma$ error with which the given ratio could be measured
for $\leff=80\fbi$ at $\rts=1\tev$ {\it assuming that the \DM\
scenario is the correct one}.
}
\begin{center}
{\footnotesize
\begin{tabular}{|c|c|c|c|c|c|}
\hline
Ratio & \DP\ & \NSM\ & \NSP\ & \HSM\ & \HSP\ \\
\hline
\multicolumn{6}{|c|} {$\langle\hh,\ha\rangle$} \\
\hline
$ [0\ell][\geq0 j]/b\anti b,t\anti t$ 
 & 12878 & 1277 & 25243 & 0.77 & 10331 \\
$ [1\ell][\geq0 j]/b\anti b,t\anti t$ 
  & 13081 & 2.41 & 5130 & 3.6 & 4783 \\
$ [2\ell][\geq0j]/b\anti b,t\anti t$ 
  & 4543 & 5.12 & 92395 & 26.6 & 116 \\
$ \hl\hl/ b\anti b$  & 109 & 1130 & 1516 & 10.2 & 6.2 \\
\hline
\multicolumn{6}{|c|} {$\hp$} \\
\hline
$ [0\ell][\geq0j]/t\anti b$ 
 & 12.2 & 36.5 & 43.2 & 0.04 & 0.2 \\
$ [1\ell][\geq0j]/t\anti b$ 
 & 1.5 & 0.3 & 0.1 & 5.6 & 0.06  \\
$\hl W/ t\anti b$ 
 & 0.8 & 0.5 & 3.6 & 7.3 & 0.3 \\
$\tau\nu/ t\anti b$ 
 & 43.7 & 41.5 & 47.7 & 13.7 & 35.5 \\
\hline
$\sum_i\Delta\chi^2_i$ & 30669 & 2493 & 124379 & 68 & 15272 \\
\hline
\end{tabular}
}
\end{center}
\label{chisqtable}
\end{table}

In Table~\ref{chisqtable} we quantify the process of excluding
the \DP, \NSM, \NSP, \HSM, and \HSP\ scenarios relative to the input
\DM\ scenario.  There we give the contribution to $\Delta\chi^2$
(computed relative to the assumed-to-be-correct 
\DM\ scenario) for each of a selection
of independently measurable ratios. Also given for each
of the incorrect scenarios is the sum of these contributions.
This table shows that the \DM\ scenario can be distinguished
from the \DP, \NSM, \NSP,  and \HSP\ scenarios
at an extremely high statistical level.
Further, even though no one of the branching fraction ratios
provides an absolutely clear discrimination between the
\DM\ and the \HSM\ scenarios, the accumulated discrimination power
obtained by considering all the ratios is very substantial. In particular,
although the ratios of Eq.~(\ref{hhhlhl}), (\ref{hazhl}),
and (\ref{hpwhl}) are only poorly measured for $\leff=80\fbi$, 
their accumulated $\Delta\chi^2$ weight can be an important
component in determining the likelihood of a given model and thereby
ruling out incorrect model choices.

Thus, consistency of all the ratios with one another
and with the measured $\mha$, neutralino and chargino
masses will generally restrict the allowed models
to ones that are very closely related.  
The likelihood or probability associated with the best fit to all these 
observables in a model that differs significantly from
the correct model would be very small.  

An important issue is the extent to which one can be
sensitive to the branching fractions
for different types of SUSY decays of the Higgs bosons,
relative to one another and relative to the overall SUSY decay
branching fraction. Rates in different channels
depend in a rather detailed fashion
upon the SUSY parameters and would provide valuable information
regarding the SUSY scenario. For example, 
in going from NS to D to HS the masses of the sneutrinos
and sleptons increase relative to those for the charginos and neutralinos.
The $\hh,\ha\to \slep^+\slep^-$ and $\hpm\to \slep^{\pm}\snu$
branching fractions should decline in comparison to 
$\hh,\ha\to\cpone\cmone$ and $\hpm\to \cpmone\cnone$, respectively.
In small sections of the D and NS scenario parameter
spaces, the sleptons and sneutrinos
are sufficiently light that $\cpmone$ decays
almost exclusively to $\slep^{\pm}\snu$ followed by 
$\slep^{\pm}\snu\to \ell^{\pm}\cnone\nu\cnone$, implying that
$\cpmone$ decays would mainly yield leptons and not jets.

The difficulty is that several different SUSY channels can contribute
to any given final state. For example, the $\ell^+\ell^-+\etmiss$
channel receives contributions from both 
$\hh,\ha\to \slep^+\slep^-$ and $\cpone\cmone$
decays; and the $\ell^{\pm}+\etmiss$ channel receives contributions
from $\hpm\to \slep^{\pm}\snu$ and $\cpmone\cnone$.  Another example,
is the purely invisible
$\hh$ or $\ha$ final state; it can arise from either 
$\cnone\cnone$ or $\snu\snu$ (with $\snu\to \nu \cnone$) production. Thus,
the physically distinct channels,
defined by the number of leptons and jets 
present,\footnote{The totally invisible final state would be 
$[0\ell][0j]$, and so forth.} typically have
multiple sources.   Still, a comparison between the rates
for the final states so-defined might be quite revealing.
For instance, if $\cpmone\to \slep^{\pm}\snu$
is not kinematically allowed, the $\cpone\cmone$ final states are expected 
to yield more $1\ell +2j$ and $0\ell+4j$
events than $2\ell+0j$ events, whereas $\slep^+\slep^-$
events will yield only $2\ell+0j$ events. Further, the $\ell$'s
must be of the same type in this latter case.
The effective branching fraction
for $\cpone\cmone\to\ell^+\ell^-+\etmiss$ with both $\ell$'s
of the same type is only 1/81.
In addition, the $\ell$'s in the latter derive from three-body decays
of the $\cpmone$, and would be much softer on average than $\ell$'s
from $\slep^+\slep^-$. Even if this difference is difficult
to see directly via distributions, it will lead to higher
efficiency for picking up the $\slep^+\slep^-$ events.

Based on the above discussion, the following ratios would appear
to be potentially useful. 
\begin{eqnarray}
&{\br(\hh\to b\anti b)\br(\ha\to [0\ell][0j])
 +\br(\ha\to b\anti b)\br(\hh\to [0\ell][0j])
\over 
  \br(\hh\to b\anti b)\br(\ha\to {\rm SUSY})
 +\br(\ha\to b\anti b)\br(\hh\to {\rm SUSY})
}  & \label{0l0j} \\
&{\br(\hh\to b\anti b)\br(\ha\to [2\ell][0j])
 +\br(\ha\to b\anti b)\br(\hh\to [2\ell][0j])
\over 
  \br(\hh\to b\anti b)\br(\ha\to {\rm SUSY})
 +\br(\ha\to b\anti b)\br(\hh\to {\rm SUSY})
}  & \label{2l0j} \\
&{\br(\hh\to b\anti b)\br(\ha\to [\geq 0\ell][0j])
 +\br(\ha\to b\anti b)\br(\hh\to [\geq 0\ell][0j])
\over 
  \br(\hh\to b\anti b)\br(\ha\to {\rm SUSY})
 +\br(\ha\to b\anti b)\br(\hh\to {\rm SUSY})
}  & \label{geq0l0j} \\
&{\br(\hh\to b\anti b)\br(\ha\to [0\ell][\geq 1j])
 +\br(\ha\to b\anti b)\br(\hh\to [0\ell][\geq 1j])
\over 
  \br(\hh\to b\anti b)\br(\ha\to {\rm SUSY})
 +\br(\ha\to b\anti b)\br(\hh\to {\rm SUSY})
}  & \label{0lgeq1j} \\
&{\br(\hh\to b\anti b)\br(\ha\to [1\ell][\geq 1j])
 +\br(\ha\to b\anti b)\br(\hh\to [1\ell][\geq 1j])
\over 
  \br(\hh\to b\anti b)\br(\ha\to {\rm SUSY})
 +\br(\ha\to b\anti b)\br(\hh\to {\rm SUSY})
}  & \label{1lgeq1j} \\
%
& {\br(\hp \to [1\ell][0j])\br(\hm\to b\anti t)
  +\br(\hm\to[1\ell][0j])\br(\hp\to t\anti b)
\over 
\br(\hp\to {\rm SUSY})\br(\hm\to b\anti t)+
\br(\hm\to {\rm SUSY})\br(\hp\to t\anti b)} 
  & \label{1l0jhp} \\
& {\br(\hp\to [\geq1\ell][0j])\br(\hm \to b\anti t)
  +\br(\hm\to [\geq1\ell][0j])\br(\hp \to t\anti b)
\over 
\br(\hp\to {\rm SUSY})\br(\hm\to b\anti t)+
\br(\hm\to {\rm SUSY})\br(\hp\to t\anti b)}
  & \label{geq1l0jhp} \\
& {\br(\hp\to [0\ell][\geq1 j])\br(\hm \to b\anti t)
  +\br(\hm\to [0\ell][\geq1 j])\br(\hp \to t\anti b)
\over 
\br(\hp\to {\rm SUSY})\br(\hm\to b\anti t)+
\br(\hm\to {\rm SUSY})\br(\hp\to t\anti b)}
.  & \label{0lgeq1jhp} 
\end{eqnarray}

Also of interest are ratios of the different numerator terms to one
another within the above neutral and charged Higgs boson sets.
All the ratios that one can form
have the potential to provide important tests of
the Higgs decays to the supersymmetric particle pair final states. 
We find that
the ratios of rates of the various SUSY channels can contribute significantly
to our ability to discriminate between different GUT scenarios.
To illustrate, we follow the same procedure as in Table~\ref{chisqtable}.
Taking $\mha=349.7\gev$ and $\mcpmone=149.5\gev$, we assume that
the correct scenario is \DM\ and compute the $\Delta\chi^2$
by which the prediction for a given ratio in the
other scenarios deviates from the \DM\ prediction.  Statistics
are computed on the basis of the expected \DM\ rates.
The resulting $\Delta \chi^2$ values
are given in Table~\ref{susychisqtable}.  Since these ratios are not
all statistically independent of one another, we do not sum their
$\Delta\chi^2_i$'s to obtain an overall discrimination level.  However,
a rough indication of the level at which 
any given scenario can be ruled out relative to the \DM\ is obtained if we
add the largest $\Delta\chi^2_i$ from the neutral Higgs
list and the largest from the charged Higgs list. The weakest discrimination
level following this procedure is $\Delta\chi^2\sim 15$ 
in the case of the \DP\ scenario. Note that this scenario is highly
unlikely on the basis of the earlier $\sum_i\Delta\chi^2_i$ value
listed in Table~\ref{chisqtable}.  In Table~\ref{chisqtable}, the weakest
discrimination was that for the \HSM\ scenario with 
$\sum_i\Delta\chi^2_i\sim 68$.  We observe from Table~\ref{susychisqtable}
that the ratio 
$\br(\hh,\ha\to [0\ell][0j])/\br(\hh,\ha\to [2\ell][0j])$
has $\Delta\chi^2_i\sim 928$ for the \HSM\ case, which would certainly
rule it out.

\begin{table}[hbt]
\caption[fake]{We tabulate $\Delta\chi^2_i$
(relative to the \DM\ scenario) for the indicated
ratios as a function of scenario,
assuming the measured $\mha$ and $\mcpmone$ values are $349.7\gev$
and $149.5\gev$, respectively. The SUSY channels have been resolved into 
final states involving a restricted number of leptons and jets.
Only those ratios with substantial power for discriminating between
scenarios are tabulated.
The error used in calculating each $\Delta\chi^2_i$ is the approximate
$1\sigma$ error with which the given ratio could be measured
for $\leff=80\fbi$ at $\rts=1\tev$ {\it assuming that the \DM\
scenario is the correct one}. 
}
\begin{center}
{\footnotesize
\begin{tabular}{|c|c|c|c|c|c|}
\hline
Ratio & \DP\ & \NSM\ & \NSP\ & \HSM\ & \HSP\ \\
\hline    
\multicolumn{6}{|c|}{$\langle\hh,\ha\rangle$} \\
\hline
$[0\ell][0j]/ {\rm SUSY}$ 
 & 3.5 & 193 & 3.4 & 1.4 & 0.6 \\
$[\geq0\ell][0j]/ {\rm SUSY}$ 
 & 0.4 & 15.3 & 6.8 & 20.9 & 15.8 \\
$[0\ell][0j]/[2\ell][0j]$ 
 & 9.6 & 503 & 0.1 & 928 & 105 \\
$[0\ell][0j]/[\geq0\ell][0j]$ 
 & 5.8 & 41.9 & 0.03 & 48.4 & 24.5 \\
$[0\ell][0j]/[0\ell][\geq1j]$ 
 & 1.4 & 1074 & 6.4 & 3.5 & 2.7 \\
$[0\ell][0j]/[1\ell][\geq1j]$ 
 & 0.3 & 3520 & 4.3 & 0 & 1.4 \\
\hline
\multicolumn{6}{|c|}{$\hp$} \\
\hline
$[\geq1\ell][0j]/ {\rm SUSY}$ 
 & 1.0 & 56.2 & 75.2 & 3.4 & 0.5 \\
$[0\ell][\geq1j]/ {\rm SUSY}$ 
 & 2.1 & 21.7 & 33.4 & 1.3 & 0 \\
$[\geq1\ell][0j]/[0\ell][\geq1j]$ 
 & 5.2 & 930 & 5738 & 4.0 & 0.4 \\
\hline
\end{tabular}
}
\end{center}
\label{susychisqtable}
\end{table}

The above illustrations demonstrate that
the ratios of rates for individual SUSY channels
correlate strongly with the underlying physics of the different GUT scenarios
(light vs. heavy sleptons in particular) and add a powerful component to our
ability to determine the correct scenario.

\section{Discussion and Conclusions}

Once the Higgs bosons are detected and their masses determined, the relative
branching fractions for the decay of a single Higgs boson
can be measured by `tagging' (\ie\ identifying) one member of the $\hh\ha$
or $\hp\hm$ pair in an all-jet mode, and then looking
at the ratios of the numbers of events in different event classes
on the opposing side.  In this way, the relative branching
ratios of Eqs.~(\ref{hhhasusy})-(\ref{hazhl}),
Eqs.~(\ref{hphmsusy})-(\ref{hpwhl}), Eqs.~(\ref{0l0j})-(\ref{1lgeq1j}),
and Eqs.~(\ref{1l0jhp})-(\ref{0lgeq1j})
can be measured with reasonable accuracy whenever parameters are such
that the final states in the numerator and denominator both
have significant event rate.
\footnote{In some cases, 
absolute event rates are so different that they would
also provide substantial discrimination
between different models, despite the possibly large
systematic errors.}
We find that the measured Higgs masses and relative branching fractions,
in combination with direct measurements of the chargino and neutralino
masses, will over-constrain and very strongly limit the possible
SUSY GUT models.  

The specific SUSY GUT models we considered are moderately
conservative in that they
are characterized by universal boundary conditions. 
The strategy for checking the consistency of a given GUT hypothesis
is straightforward. First, the measured $\ha$, neutralino
and chargino masses are, in almost all cases,
already sufficient to determine the $\mhalf$
and $\tanb$ values required in the given GUT scenario with good precision.  
The Higgs sector branching fractions
can then be predicted and become an
important testing ground for the consistency of the proposed GUT hypothesis
as well as for testing the MSSM two-doublet Higgs sector structure per se.
Typically, a unique model
among the six rather similar models is singled out by combining
measurements from the Higgs sector with those from conventional SUSY
pair production. In short, measurements deriving from pair production
of Higgs particles can have a great impact
upon our ability to experimentally determine the correct SUSY GUT model.

The above discussion has left aside the fact that
for universal soft-scalar masses the measured value
of the slepton mass would determine
the relative magnitude of $m_0$ and $\mhalf$, thereby restricting
the possible scenarios (see Table~\ref{mhalftanbtable}). 
However, if the soft-scalar
slepton mass is not the same as the soft-scalar Higgs field masses
at the GUT scale, the branching fraction ratios would give the best
indication of the relative size of the soft-scalar Higgs mass
as compared to $\mhalf$.

More information regarding the slepton/sneutrino mass scale
and additional ability to discrminate between models
are both realized by subdividing the SUSY decays of the Higgs bosons 
in a way that is sensitive to the relative branching fractions for
slepton/sneutrino vs. chargino/neutralino decays. Slepton/sneutrino
channels essentially only produce leptons in the final state, whereas
the jet component is typically larger than
the leptonic component for chargino/neutralino decays (other than
the totally invisible $\cnone\cnone$ mode). Thus, we 
are able to define individual
SUSY channels, characterized by a certain number of leptons
and/or jets, which display a strong correlation
with the slepton/sneutrino decay component. We find that these
individual channels have sufficiently large event rates
that the ratios of the branching fractions for these channels can
typically be determined with reasonable statistical precision. 
Excellent discrimination between models on this basis is found.

In conclusion, our study shows that not only will detection
of Higgs pair production in $\epem$ or $\mupmum$
collisions (at planned luminosities) be possible for most of the kinematically accessible
portion of parameter space in a typical GUT model, but also the detailed rates
for and ratios of different neutral and charged Higgs decay final states
will very strongly constrain the choice of GUT-scale boundary conditions.
In estimating experimental sensitivity for Higgs pair detection
and for measuring Higgs masses and branching fractions, we included
substantial inefficiencies and all relevant branching fractions.
Although we believe that our estimates are relatively conservative,
it will be important to re-visit this analysis using a full Monte Carlo
detector simulation. 

%

\end{document}